\documentclass[aps,prd,10pt,twocolumn,superscriptaddress,showpacs,nobibnotes,longbibliography]{revtex4-1}

\usepackage{gensymb}
\usepackage{hyperref}
\usepackage{graphicx}
\usepackage{amsfonts,amsmath,amssymb,bm,bbm}
\usepackage{color,soul}
\usepackage{slashed}
\usepackage[toc,page]{appendix}

\hypersetup{
    pdfnewwindow=true,      
    colorlinks=true,       
    linkcolor=blue,          
    citecolor=blue,        
    filecolor=blue,      
    urlcolor=blue        
}

\begin{document}

\title{Strong New Limits on Light Dark Matter from Neutrino Experiments}

\author{Christopher V. Cappiello}
\email{cappiello.7@osu.edu}
\thanks{\scriptsize \!\! \href{http://orcid.org/0000-0002-7466-9634}{orcid.org/0000-0002-7466-9634}}
\affiliation{Center for Cosmology and AstroParticle Physics (CCAPP), Ohio State University, Columbus, OH 43210}
\affiliation{Department of Physics, Ohio State University, Columbus, OH 43210}

\author{John F. Beacom}
\email{beacom.7@osu.edu}
\thanks{\scriptsize \!\!  \href{http://orcid.org/0000-0002-0005-2631}{orcid.org/0000-0002-0005-2631}}
\affiliation{Center for Cosmology and AstroParticle Physics (CCAPP), Ohio State University, Columbus, OH 43210}
\affiliation{Department of Physics, Ohio State University, Columbus, OH 43210}
\affiliation{Department of Astronomy, Ohio State University, Columbus, OH 43210} 

\date{November 18, 2019}

\begin{abstract}
The non-detection of GeV-scale WIMPs has led to increased interest in more general candidates, including sub-GeV dark matter. Direct-detection experiments, despite their high sensitivity to WIMPs, are largely blind to sub-GeV dark matter. Recent work has shown that cosmic-ray elastic scattering with sub-GeV dark matter would both alter the observed cosmic ray spectra and produce a flux of relativistic dark matter, which would be detectable with traditional dark matter experiments as well as larger, higher-threshold detectors for neutrinos. Using data, detectors, and analysis techniques not previously considered, we substantially increase the regions of parameter space excluded by neutrino experiments for both dark matter-nucleon and dark matter-electron elastic scattering. We also show how to further improve sensitivity to light dark matter.

\end{abstract}

\maketitle

% % % % % % % % % % % % % % % % % % % % % % % % % % % % % % % % % % % % % % INTRODUCTION % % 

\section{Introduction}

Though it makes up most of the mass in the universe, dark matter (DM) is only known to interact gravitationally. As a result, its particle mass and scattering cross sections are unknown. Direct-detection experiments, collider searches, and a wide array of cosmological and astrophysical studies have searched for signs of DM interacting with either nucleons or electrons with no clear signals to date \cite{dd7, dd8, dd9, dd6, Kahrev, ATLAS17, CMS18, Lea18}. 

%Such searches have often focused on GeV-scale WIMPs (weakly interacting massive particles), and direct detection experiments in particular set strong limits on DM scattering over a wide mass range, but rapidly lose sensitivity for masses below $\sim 1$ GeV. Collider experiments have searched for WIMPs by considering missing transverse energy in collisions, and astrophysical studies have searched for high-energy particles produced by WIMP annihilation and decay. Although WIMPs are nowhere near being ruled out \cite{Lea18}, their non-detection at dedicated experiments has led to increased interest in other candidates, including sub-GeV DM \cite{Ess12, Ess15, Ali15, Ang17, An18, Emk17, ADMX18}.

%For masses below 1 GeV, the experimental sensitivity is much worse. 

Direct detection experiments set strong limits on DM heavier than $\sim 1$ GeV, but rapidly lose sensitivity below this point, because DM that is too light doesn't carry enough momentum to trigger them. This means different approaches are needed to search for light DM. Cosmology and astrophysics set limits on the DM-nucleon cross section for low masses, but these are of order $\sim10^{-28}\,$cm$^2$, more than fifteen orders of magnitude weaker than direct detection constraints on GeV-scale DM \cite{Nad19, Bod18, Glu18, Sla18, Xu18, Bho18a, Bho18b, Cap19, Wad19}. Colliders probe much smaller cross sections, but these limits are more model dependent, and more importantly, there is a ceiling above which DM would interact in the detectors, making traditional missing-energy searches insensitive \cite{Bai11, Coh15, Dac15}. Additional, model-dependent limits may be set based on energy deposition in white dwarfs or neutron stars \cite{Bar17, Raj17, Das19}, Big Bang Nucleosynthesis \cite{Krn19, Bon19}, DM production in cosmic ray (CR) showers \cite{Alv19}, the cooling of stars and supernovae \cite{Raf99, Fay06, Pos08, Raf08, Guh18}, and Casimir-Polder type forces \cite{Fic18, Bra18}. There is a large gap between collider and astrophysical/cosmological limits that needs to be probed; see Fig. 1 of Ref.~\cite{Cap19}.

The DM-electron elastic scattering cross section can be probed by direct detection experiments for masses well below a GeV, but these experiments reach a similar kinematic limit around 1 MeV. Cosmological and astrophysical limits on DM-electron scattering exclude cross sections above $\sim10^{-28}\,$cm$^2$ \cite{Ali15, Cap19, Wad19} (but neutron star heating may provide much stronger constraints \cite{Bel19}). And collider limits (e.g., Refs.~\cite{Fox11, Ess13}) should have a ceiling analogous to the one in the nucleon case, though as far as we know, it has not been calculated. For the electron case as well, new ideas are needed to close the window between cosmological and collider limits.

\begin{figure}[t]
\centering
\includegraphics[width=\columnwidth]{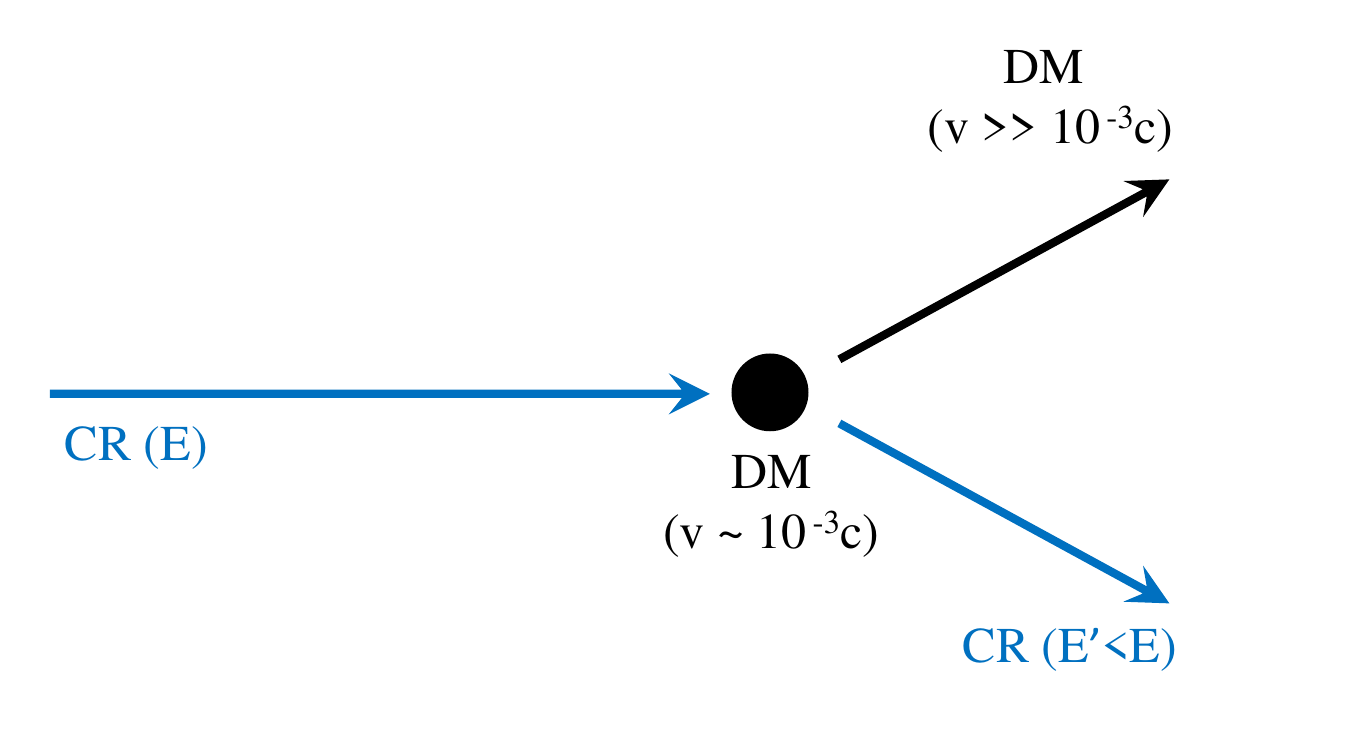}
\caption{Schematic diagram of CR-DM scattering. A CR with velocity much higher than that of the DM ($\sim10^{-3} c$) scatters with DM, transferring some of the CR's kinetic energy and boosting the DM to higher velocity.} 
\label{fig:diagram}
\vspace{-0.25 cm}
\end{figure}

Recently, we showed that for allowed values of the DM-proton and DM-electron elastic scattering cross sections, CRs would lose enough energy in collisions with light DM to alter their observed spectra, and used this observation to set strong limits on these cross sections \cite{Cap19}. Figure~\ref{fig:diagram} shows a schematic diagram of this scattering process. Following our paper, Refs.~\cite{Bri18} and~\cite{Ema18} considered the complementary effect: collisions with CRs can upscatter DM particles to relativistic or near-relativistic speeds, allowing DM particles much lighter than a GeV to produce detectable recoils even in relatively high-threshold detectors. Ref.~\cite{Bri18} used data from MiniBooNE, Borexino, and Xenon1T to derive large new exclusion regions on sub-GeV DM scattering with nucleons; similarly, Ref.~\cite{Ema18} used data from MiniBooNE and Super-Kamiokande (Super-K) to set strong new constraints on sub-MeV DM scattering with electrons. (These exclusion regions are complementary to those recently derived based on solar reflection \cite{An18, Emk17}; see below.) We emphasize that DM upscattering by CRs is very different from the scenario usually referred to as ``boosted" DM, in which energetic DM particles of one species are produced by pair annihilation of a second, heavier DM species \cite{Aga14, Cha18, Ha18, McK18}.

Here we expand on the results of Refs.~\cite{Bri18} and \cite{Ema18} by considering detectors and data sets that these papers did not use, and analyzing this additional data more precisely. We use our own numerical code to more carefully model attenuation of the DM flux by nuclei in Earth's crust. For DM-nucleon scattering, we consider data from Daya Bay, KamLAND, and PROSPECT, and for DM-electron scattering, we consider lower-energy Super-K data than considered in Ref.~\cite{Ema18}. 

\section{From Cosmic Ray Spectra to Recoil Distributions}\label{sec:recoildist}

In this section, we introduce the basics of CR propagation and compute the recoil spectrum seen in a detector for a given CR spectrum. This overview sets the foundation for our new results in subsequent sections.

\subsection{Cosmic Ray Inputs}

CRs are energetic charged particles, mostly protons but also including electrons and a range of heavier nuclei, which, at the energies we consider, are thought to largely be accelerated in supernova remnants \cite{Bla11, Bla13, Mor17, Byk18}. CRs are energetic enough to upscatter DM particles to relativistic speeds; the observed proton and electron CR spectra at Earth both peak at around 1 GeV, ultra-relativistic energy for electrons and moderately relativistic for protons, and extend many orders of magnitude higher in energy.

CRs are trapped by galactic magnetic fields in a thick, disklike halo around the galactic disk for far longer than it would take to cross the galaxy in a straight line, and their propagation within this halo is described by a diffusion equation. The size of the halo is uncertain, but we adopt the relatively conservative geometry used in Ref.~\cite{Ema18}, assuming CRs are uniformly distributed in a cylinder with radius $R = 10$ kpc and half-height $h = 1$ kpc; considering a larger volume, which is probably more realistic (e.g., Ref.~\cite{Str00}), would produce stronger limits because the upscattered DM flux would be larger.

The interstellar CR spectrum is different from the spectrum observed at Earth due to solar modulation. For energies below a few GeV, solar modulation suppresses the CR flux at Earth, contributing to the peak mentioned above. Voyager 1 has measured the local interstellar spectra (LIS) of CR electrons and various nuclei down to energies of 1--10 MeV \cite{Cu16}, and recent papers have computed the LIS down to the lowest energies of the Voyager electron, proton, and helium spectra \cite{Bos17, Bos18a, Bos18b}. The LIS has been shown to agree well with the CR spectra elsewhere in the galaxy as inferred by gamma-ray observations, with the CR density increasing somewhat at smaller galactic radii \cite{Ack11, Yan16, Aha18}; we therefore use for the galactic CR spectra the LIS computed in Refs.~\cite{Bos17, Bos18a}. In constraining DM scattering with nuclei, we restrict our attention to the CR proton and helium spectra.

\subsection{CR-DM Scattering}

Following Refs.~\cite{Cap19, Bri18, Ema18}, we assume that both the DM-electron and DM-nucleon cross sections are energy independent, and that DM-electron scattering is isotropic in the center-of-momentum (CM) frame, while DM-nucleus scattering deviates from isotropy only due to nuclear form factors. The assumption of energy independence is a simplification that allows for straightforward comparison between our results and constraints from astrophysics, cosmology, and traditional direct detection. In Sec.~\ref{sec:discussion}, we discuss the implications of this choice and possible extensions, such as considering inelastic scattering with nuclei. Following Ref.~\cite{Bri18}, we expect that the elastic scattering cross section, enhanced by a factor of $A^2$, is dominant in the DM energy range we consider.

Typical DM velocities in the galaxy ($\sim10^{-3} c$) are small compared to the velocities of the CRs we consider, so we treat the DM as being at rest. The kinetic energy transferred to a stationary DM particle of mass $m_{\chi}$ by a CR with mass $m_{\rm CR}$ and kinetic energy $T_{\rm CR}$ is

\begin{equation}
T_{\chi} = \frac{T_{\rm CR}^2+2m_{\rm CR}T_{\rm CR}}{T_{\rm CR}+(m_{\rm CR}+m_{\chi})^2/(2m_{\chi})}\left(\frac{1-\cos \theta}{2}\right)\, ,
\end{equation}
where $\theta$ is the CM scattering angle. Consequently, the maximum recoil energy is

\begin{equation}\label{max}
T_{\chi}^{\rm max} = \frac{T_{\rm CR}^2+2m_{\rm CR}T_{\rm CR}}{T_{\rm CR}+(m_{\rm CR}+m_{\chi})^2/(2m_{\chi})}\, .
\end{equation}
Inverting this equation gives the minimum CR energy, $T_{\rm CR}^{\rm min}(T_{\chi})$ required to produce a DM recoil energy $T_{\chi}$:

\begin{equation}\label{min}
T_{\rm CR}^{\rm min} = \left(\frac{T_{\chi}}{2}-m_{\rm CR}\right)\left(1 \pm \sqrt{1+\frac{2T_{\chi}}{m_{\chi}}\frac{(m_{\rm CR}+m_{\chi})^2}{(2m_{\rm CR}-T_{\chi})^2}}\right)\, ,
\end{equation}
where the $+$ applies for $T_{\chi} > 2m_{\rm CR}$ and the $-$ applies for $T_{\chi} < 2m_{\rm CR}$.

\subsection{DM Flux and Spectrum}

For the DM density profile, we use a Navarro-Frenk-White (NFW) profile \cite{Nav96} with scale radius $r_s = 20$ kpc and a density at Earth of 0.3 GeV/cm$^3$. This is the same scale radius used in Ref.~\cite{Ema18}, though they use a local density of 0.42 GeV/cm$^3$. The difference in density has only a mild impact on our results, as discussed below.

The differential flux at Earth of DM upscattered by collisions with CRs of species $i$ (in terms of incident CR energy $T_i$) is given by a line-of-sight integral:

\begin{equation}\label{DMdist}
\frac{d\Phi_{\chi}}{dT_i}=\int\frac{d\Omega}{4\pi}\int_{l.o.s.}dl \, \sigma_{\chi i}\frac{\rho_{\chi}}{m_{\chi}}\frac{d\Phi_i}{dT_i}\,.
\end{equation}
We integrate over the full CR halo with the geometry given above. Note that this flux is in terms of the CR energy, not the DM energy. To convert this into a DM energy spectrum, we integrate over CR energies:

\begin{equation}\label{DMdist2}
\frac{d\Phi_{\chi}}{dT_{\chi}} = \int_0^{\infty}dT_i \frac{d\Phi_{\chi}}{dT_i} \frac{1}{T_{\chi}^{\rm max}(T_i)}\Theta[T_{\chi}^{\rm max}(T_i)-T_{\chi}]\,.
\end{equation}

\begin{figure*}[t]
\centering
\includegraphics[width=\textwidth]{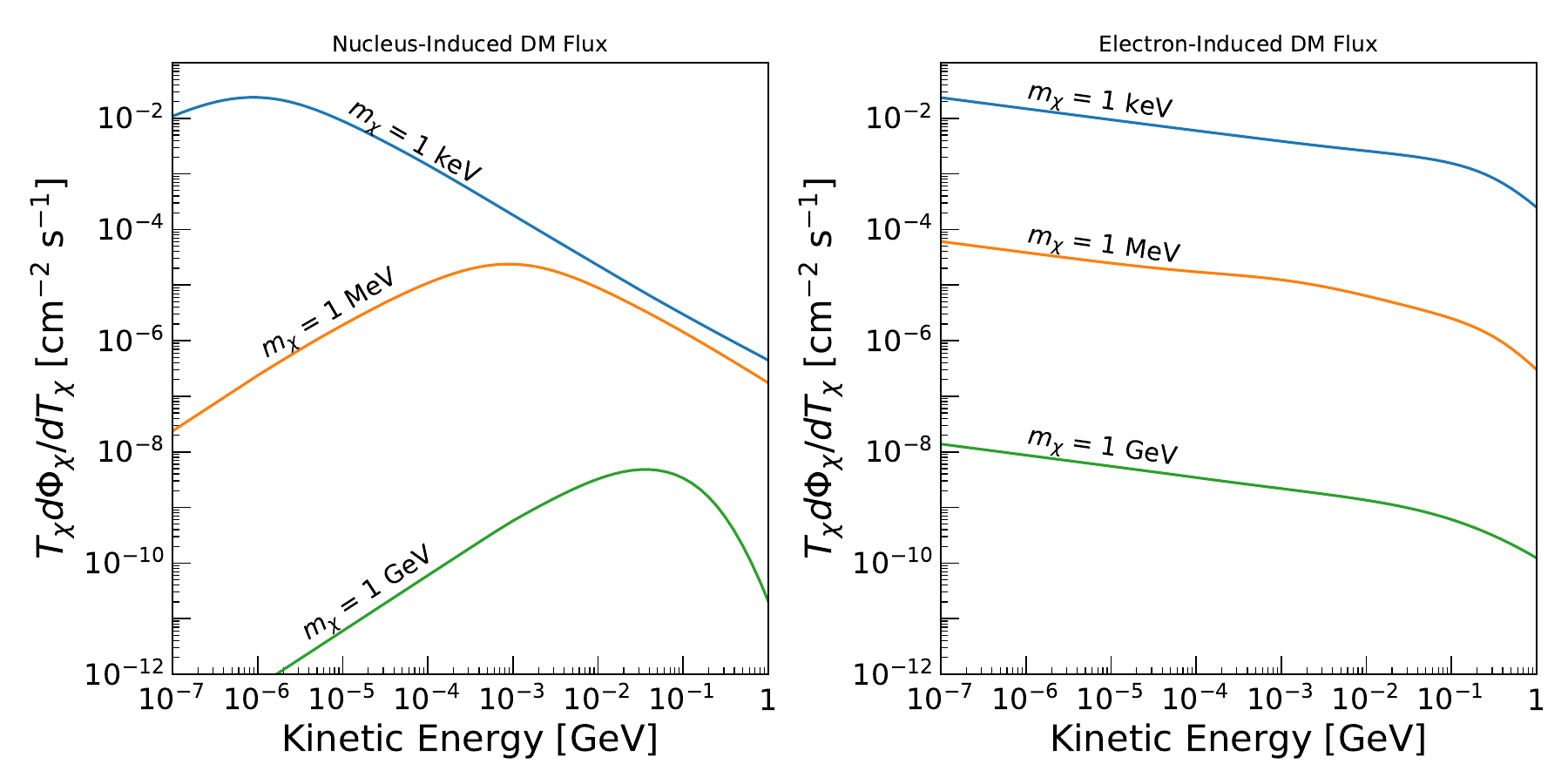}
\caption{Left: Flux of DM upscattered by CR nuclei (protons and helium), assuming a DM-nucleon elastic scattering cross section of $10^{-30}$ cm$^2$, and DM masses as labeled. Right: Flux of DM upscattered by CR electrons, assuming a DM-electron elastic scattering cross section of $10^{-30}$ cm$^2$, and the same DM masses.} 
\label{fig:dmdist}
\vspace{-0.25 cm}
\end{figure*}

Figure~\ref{fig:dmdist} shows the DM spectra reaching Earth after collisions with either protons (plus helium) or electrons, for several masses. For $m_{\chi} \ll m_{\rm CR}$, the proton-induced spectra show breaks at $T_{\chi} \simeq m_{\chi}$; in the limit where $m_{\rm CR} > m_{\chi}$ and $m_{\rm CR} > T_{\chi}$, the term in the square root in Eq.~(\ref{min}) is $1 + T_{\chi}/(2m_{\chi})$, leading to the observed break. For $m_{\chi} \simeq 1$ GeV, the proton and helium form factors begin to matter, causing the break to be slightly lower in energy than $m_{\chi}$. For the electron-induced spectra, because $m_{\rm CR}$ is not necessarily greater than $T_{\chi}$, the corresponding break is roughly compensated by an additional break coming from the factors of $2m_{\rm CR} - T_{\chi}$, which for protons showed up at too high energy to be relevant. The break that does appear is due to the break in the electron LIS, which was included in order to fit the low-energy Voyager data. It is also interesting to note that for light DM, the electron-induced flux is higher at high energy than the proton-induced flux, despite the proton CR flux being higher than that of electrons. This is due to electrons transferring a larger fraction of their energy to light DM than protons because they are closer in mass to the DM.

The incoming DM flux should have significant directional variation: the highest flux should come from the direction of the galactic center, where the line-of-sight integrated DM density is highest. However, scintillator detectors (and typical DM detectors) lack the directional sensitivity to use this information. We discuss below how Super-K's directional sensitivity could be useful for improving constraints on DM-electron scattering. 

\subsection{Attenuation of the DM Flux}

Direct-detection experiments are blind to DM with sufficiently large cross sections, as it would be stopped from reaching the detectors by scattering in the atmosphere, Earth, and detector shielding. This effect is often neglected for direct detection experiments, as they are designed to probe such low cross sections that Earth is effectively transparent to DM. However, for sub-GeV DM, existing limits are weak enough that for the cross sections we hope to probe, attenuation may be significant.

References~\cite{Bri18, Ema18, Dent19} modeled attenuation using a ballistic-trajectory approach that assumes that DM travels in a straight line from the top of the atmosphere to the detector, losing energy as it scatters but not changing direction. This approach is reasonable for DM much heavier than the target particles (as in Ref.~\cite{Kav17}), but for light DM ($m_{\chi} \lesssim m_A$), it is clearly not well motivated: particles may backscatter into the atmosphere, and even DM that eventually reaches the detector may have a much longer path length than the straight-line distance to the detector \cite{Mah17}. On the other hand, Ref.~\cite{Bri18} neglects the form factor of nuclei in Earth's crust, which should make their results more conservative. While numerical codes have been used to study attenuation of nonrelativistic DM \cite{Mah17, Mah17b, Emk17c, Emk18}, no such code has previously been used to study CR-upscattered DM.

To model propagation of relativistic or nearly relativistic DM, we wrote our own propagation code, modeled somewhat on the publicly available DMATIS code \cite{Mah17, DMATIS} but including relativistic kinematics and the kinetic energy distribution of CR-upscattered DM. Our purpose here is not to model attenuation in a detector-independent way, but rather to compute the recoil spectrum in a particular detector, which can then be compared to data. In the next section, we describe our code and compare its results with those of the ballistic approximation.

\subsection{Target Recoil Distribution}

Denoting the DM flux at detector depth $z$ as $d\Phi_{\chi}^z/dT_{\chi}$, the differential recoil rate per target particle $k$ (nucleus or electron) is then

\begin{equation}\label{recoil}
\frac{d\Gamma}{dT_k} = \sigma_{\chi k} \int_{T_{\chi}^{\rm min}}^{\infty}dT_{\chi}\,\frac{1}{T_k^{\rm max}}\frac{d\Phi_{\chi}^z}{dT_{\chi}}\,,
\end{equation}
where $T_k^{\rm max}$ and $T_{\chi}^{\rm min}$ are obtained from Eqs.~(\ref{max}) and (\ref{min}) respectively, by replacing $\chi \rightarrow k$ and CR $\rightarrow \chi$.

It is important to note that because $d\Phi_{\chi}^z/dT_{\chi}$ contains a factor of the DM-proton or DM-electron cross section, the event rate scales as cross section squared: one factor of the cross section from the DM being struck by a CR, and one factor from the DM interacting in the detector. As discussed below, this makes our final results fairly insensitive to astrophysical uncertainties because they scale only with the square root of astrophysical inputs.

\section{Propagation of Dark Matter to the Detector}

The cross sections we probe are large enough that dark matter may be significantly slowed or deflected by Earth's crust and atmosphere before reaching a detector. Unlike previous work, which employed an analytic but physically inaccurate ballistic-trajectory method to model attenuation, we develop our own numerical propagation code, using this code to treat attenuation and comparing its results with the ballistic model.

\subsection{Ballistic-Trajectory Approximation}

As discussed above, studies of DM attenuation in Earth often use a simple, ballistic-trajectory approximation to model the DM energy loss in the crust, atmosphere, and/or detector shielding \cite{Sta90, Kou14, Bri18, Ema18, Dent19}. In this formalism, the energy loss rate (in units of MeV/cm) is
\begin{equation}
\frac{dT_{\chi}}{dx} = -\sum_j n_j \int_0^{T_r^{\rm max}}dT_r\,\frac{d\sigma_{\chi j}}{dT_r}T_r\,,
\end{equation}
where $T_r$ is the energy lost by the DM particle in a collision and the sum is over all relevant target particles (such as various nuclei). For isotropic scattering, $d\sigma_{\chi j}/dT_r = \sigma_{\chi j}/T_r^{\rm max}$, which gives

\begin{equation}\label{attenuation}
\frac{dT_{\chi}}{dx} = -\frac{1}{2}\sum_in_j\sigma_{\chi j}T_r^{\rm max}\,.
\end{equation}
For spin-independent DM-nucleus scattering, the scattering cross section $\sigma_{\chi A}$ is

\begin{equation}\label{crosssec}
\sigma_{\chi A} = \sigma_{\chi N} A^2 \left[\frac{m_A(m_{\chi}+m_N)}{m_N(m_{\chi}+m_A)}\right]^2 G_A^2(Q^2)\,,
\end{equation}
where $G_A(Q^2)$ is the nuclear form factor.

This energy loss rate is then used to compute the spectrum at depth $z$, given the spectrum at the top of the atmosphere. However, while the ballistic-trajectory approach is reasonable for sufficiently heavy DM, it is not a good qualitative or quantitative description of the propagation process for light DM. To more accurately model DM propagation through the atmosphere, crust, and detector shielding, we use our own numerical code.

\subsection{Description of Numerical Code}

Our code models the atmosphere as being composed of 21\% oxygen and 79\% nitrogen, neglecting additional elements. It models Earth's crust as being composed of oxygen, silicon, aluminum and iron, with the same abundances used in the DMATIS code \cite{DMATIS}. Any concrete shielding (relevant for PROSPECT, discussed below) is modeled as having the same composition as the crust.

Our code begins by selecting a kinetic energy from the input DM energy distribution, and initializing a DM particle at the top of the atmosphere, moving toward the detector. In each of the regions defined above, the code computes the mean free path based on the density of material, elemental abundances, and DM-nucleon cross section. It samples from the resulting path length distribution to determine the distance traveled before one interaction, and selects the nucleus encountered in the first scattering with the probability of each nucleus weighted by $A^2$ (from the scaling of the cross section) times the abundance (which scales as $1/A$ for a given mass density). It then uses the form factor for that nucleus and the DM energy to determine the scattering angle and energy loss of that collision. For elements in the crust, shielding, and atmosphere, we employ the commonly used Helm form factor \cite{Hel56, Dud06}.

This process is repeated until the particle either scatters back into the atmosphere, loses too much energy to produce an event in the energy range we consider, or reaches the depth of the detector. If it reaches the depth of the detector, the recoil energy is computed based on the DM energy and the form factor of the target nucleus. This process is repeated for $10^6$ particles, and the recoil spectrum plotted. The recoil spectrum is then multiplied after the fact by the probability of a particle actually interacting with a nucleus in the detector.

\subsection{Comparison of Numerical Method and Ballistic Approximation}

References~\cite{Mah17b, Emk18} have compared the results of numerical codes and the ballistic-trajectory approximation, and generally speaking found the ballistic-trajectory approach to be conservative for the detectors and mass ranges they considered. However, these conclusions cannot be naively applied to lower-mass, relativistic DM in higher-threshold detectors. Thus it is necessary to do our own comparison.

As an example, we compute the DM-induced proton-recoil spectrum in Daya Bay, at a mass of 1 MeV and a cross section of $5.0\times10^{-28}$ cm$^2$, using both the ballistic approximation (neglecting form factors) and our numerical code. We propagate $10^5$ DM particles to a depth of 100 m, modeling only particles with kinetic energy up to 1 GeV. We then compare the predicted spectra of DM-induced events in Daya Bay with the observed event spectrum. (In the next section, we provide details of the Daya Bay experiment and describe how we use their data to constrain DM properties.)

\begin{figure}[t]
\centering
\includegraphics[width=\columnwidth]{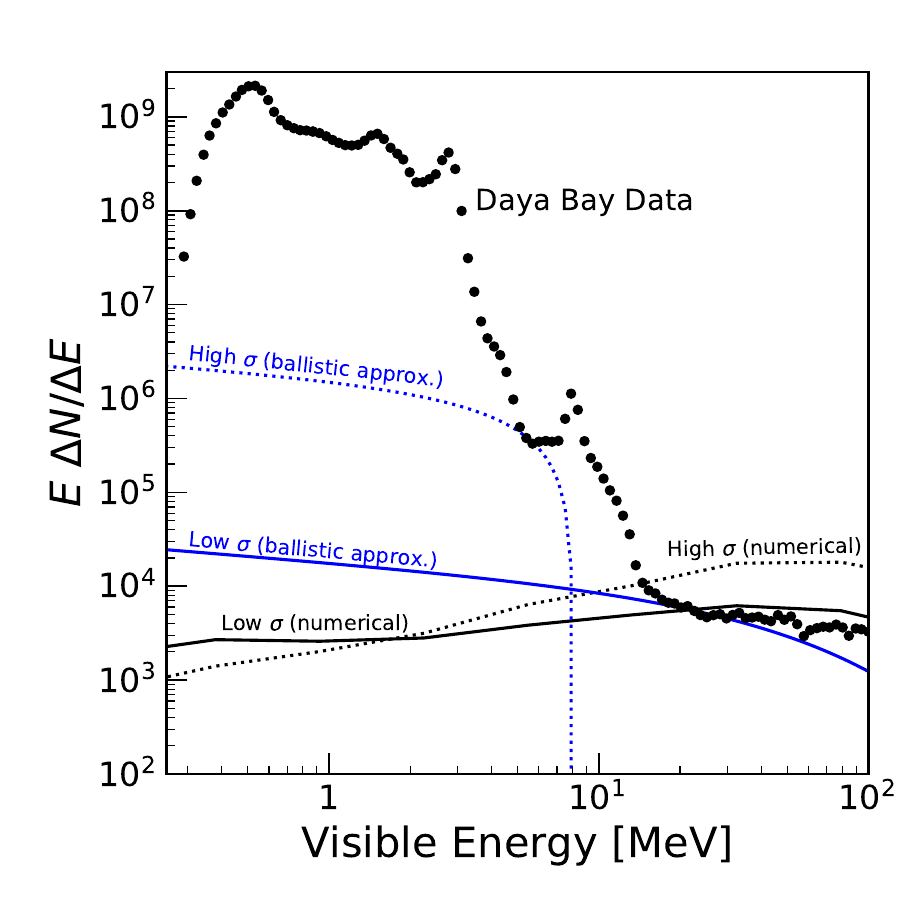}
\caption{Daya Bay background-dominated data \cite{An12} compared to predicted DM-induced spectra. Data from a combined 3.6 kton--day exposure of two Daya Bay antineutrino detectors (black points). Also shown are ballistic (blue) and numerically (black) computed spectra for a mass of 1 MeV. Dotted lines are the two calculations of the spectrum at a cross section of $5 \times 10^{-28}\, \rm{cm}^2$ (denoted ``High $\sigma$"); solid lines are the spectrum at $5 \times 10^{-29}\, \rm{cm}^2$ (denoted ``Low $\sigma$"). These cross sections are barely excluded in the ballistic case, as the DM-induced spectra are barely above the data, but much more strongly excluded in the numerical case.} 
\label{fig:dbatten}
\vspace{-0.25 cm}
\end{figure}

Figure~\ref{fig:dbatten} shows the spectrum of DM-induced events computed using the ballistic approximation (blue) and our numerical code (black). The ballistic approximation overpredicts the event rate at low energy because it neglects changes in direction during propagation. For low-energy DM particles, say with energy $\sim200$ MeV when considering MeV DM scattering with silicon, the form factor's deviation from unity is largely negligible, and particles scatter nearly isotropically. More than half of these particles are backscattered out of the atmosphere, and even the particles that reach the detector's depth scatter far more times than the ballistic trajectory approach assumes, and lose most of their energy. On the other hand, higher-energy particles (above a couple hundred MeV) are weighted much more heavily toward forward scattering and lower energy loss. Because the ballistic approximation underpredicts the DM-induced event rate where the background rate is lowest, \textit{our numerical code finds that the ceiling should be higher than would be computed using the ballistic approximation}.

The difference between the ballistic and numerical treatments of attenuation is important not just for computing the ceiling of an exclusion region, but also for determining what data to use to set that ceiling. Ref.~\cite{Dent19}, using the ballistic approximation, shows that the ceiling of their XENON1T exclusion region would move up noticeably if the threshold were reduced to 1 keV. This is almost certainly due to the high-energy cutoff in the DM spectrum seen in the analytic approximation. However, numerically computing attenuation does not produce this same cutoff, making the threshold energy less important.

\section{Constraining the DM-Nucleon Cross Section}\label{results}

In this section, we use the above formalism, along with data from Daya Bay, KamLAND, and PROSPECT, to set new limits on the DM-nucleon cross section $\sigma_{\chi N}$. As above, we consider only elastic scattering on protons in the detectors. We do not consider elastic scattering on carbon nuclei because the effect of quenching is much greater for carbon than for protons (see below), but in Sec.~\ref{sec:discussion} we discuss inelastic collisions with various nuclei.

\subsection{Our Improvements Over Previous Work}

On top of properly modeling attenuation, we build on previous work by considering additional data and detectors, allowing us to probe additional parameter space. While the exclusion regions of Ref.~\cite{Bri18} are broad in both mass and cross section, there remains a gap between the MiniBooNE and XENON1T regions, which we almost completely close. Using data from Daya Bay, we almost completely close this gap. KamLAND, the deepest detector we consider, has the lowest ceiling but also the lowest background rate, making it sensitive to cross sections much lower than MiniBooNE and comparable to XENON1T. Finally, we compute projected sensitivities for PROSPECT, whose position at the Earth's surface will make it sensitive to cross sections higher than MiniBooNE, and JUNO, a future detector with the potential to probe cross sections lower than XENON1T.

Typical direct detection experiments use heavy nuclei such as xenon to probe small cross sections, because of the cross section's commonly assumed scaling with mass number. However, this scaling need not hold in general: DM may couple differently to protons and neutrons, and either way the scaling may break down at large cross sections \cite{Dig19}. It is therefore valuable, if possible, to constrain the DM-proton cross section through scattering with individual protons. By focusing on proton recoils in scintillator detectors, we are able to probe the DM-proton cross section \textit{directly}, making our results complementary to the XENON1T limits.

We compute exclusion regions for these detectors for DM masses from 1 keV to 1 GeV. Masses below about 1 keV are disfavored because of constraints from structure formation \cite{Aba05,Vie06,Boy08}, and fermionic DM cannot be lighter than around 100 eV due to phase space constraints \cite{Boy09,DiP17}. However, our formalism is perfectly valid for masses far below 1 keV, and could thus constrain models that could evade the cosmological constraints.

\subsection{Proton Recoil Distribution}
We consider proton and helium CRs elastically scattering with DM particles, neglecting heavier nuclei, using the LIS computed by Ref.~\cite{Bos17} for rigidities from 2 MeV to 100 TeV. The cross section is given by Eq.~\ref{crosssec}, where for protons and helium form factor $G_A(Q^2)$ is
\begin{equation}
G_A(Q^2) = 1/(1+Q^2/\Lambda_A^2).
\end{equation}
For a vector current, as considered by Ref.~\cite{Bri18}, $\Lambda_p \simeq$ 770 MeV and $\Lambda_{He} \simeq$ 410 MeV \cite{Ang04}. Thus we compute the boosted DM distribution by summing over Eq.~(\ref{DMdist}) hydrogen and helium with the cross section parametrized as above, and inserting the result into Eq.~(\ref{DMdist2}). Depending on the energy, the contribution of helium to the upscattered DM flux ranges from negligible compared to the proton contribution, to roughly comparable with it; see Fig. 1 of Ref.~\cite{Bri18}.

\subsection{New Limits from Daya Bay Data}

The Daya Bay Reactor Neutrino Experiment consists of eight antineutrino detectors (ADs) divided among three experimental halls (EHs) inside a mountain, with vertical overburdens ranging from 250 to 860 meters water equivalent (m.w.e.)\ \cite{An12, Cao16}. The shallowest of these, EH1,  is shallower than XENON1T, Borexino, and KamLAND by about a factor of 10, and deeper than MiniBooNE by around the same factor. It is located near the side of the mountain's base (see Fig. 1 of Ref.~\cite{An17} for a diagram of the mountain). We consider only the EH1 ADs, and conservatively assume that all DM arriving from below the horizon is blocked by Earth. Based on the 3-d map, we conservatively assume that the mountain blocks an additional 1/4 of the above-horizon flux, and that the rest is attenuated by 250 m.w.e.\ of Earth. We make no attempt to reduce the background by a similar angular factor (for Daya Bay or any other experiment), and instead compare this DM flux to the total Daya Bay event spectrum as described below

We use our numerical code described above to model attenuation by Earth's crust. Because of the detector depth, attenuation in the atmosphere is negligible for Daya Bay and KamLAND, so we set our code to ignore attenuation in the atmosphere.

To compare the DM-induced recoil spectrum to the reported data, we write the recoil spectrum in terms of electron equivalent energy $T_e$, given by Refs.~\cite{Bea02, Das11} as

\begin{equation}
T_e(T_p) = \int_0^{T_p}\frac{dT_p}{1+k_B\langle dT_p/dx \rangle}\,,
\end{equation}
where $T_p$ is the proton recoil energy and $k_B$ is the material-dependent Birks' constant \cite{Bir51}. For Daya Bay, we use $k_B = 0.0096$ cm/MeV for linear alkylbenzene \cite{von13}. For KamLAND, we use $k_B = 0.015$ cm/MeV, as reported by Ref.~\cite{Yos10}. For PROSPECT, we assume $k_B = 0.0111$ cm/MeV, following the simpler of the two Birks' models in Ref.~\cite{Nor17}.

\begin{figure}[t]
\centering
\includegraphics[width=\columnwidth]{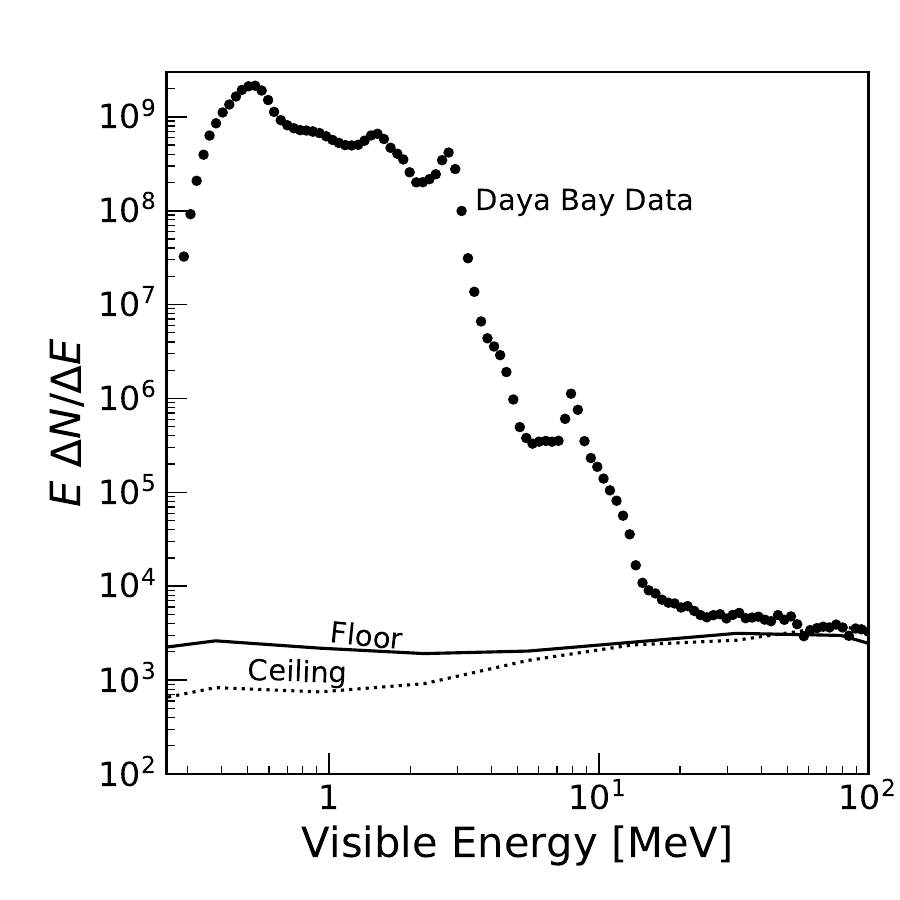}
\caption{Daya Bay background-dominated data \cite{An12} compared to predicted DM-induced spectra. Data from a combined 3.6 kton--day exposure of two Daya Bay ADs (black points). Also shown are two predicted DM-induced spectra for a mass of 1 MeV and cross sections of $6 \times 10^{-29}\, \rm{cm}^2$ and $7 \times 10^{-28}\, \rm{cm}^2$, corresponding respectively to the floor and ceiling of the Daya Bay exclusion region in Fig.~\ref{fig:protonlimit}.} 
\label{fig:dbplot}
\vspace{-0.25 cm}
\end{figure}

Ref.~\cite{An12} shows the event spectrum of the EH1 ADs after muon veto cuts have been applied, taken over three months, in the energy range from 0.3 to 100 MeVee. Here MeVee is short for MeV electron equivalent, meaning the kinetic energy reconstructed from the observed scintillation signal assuming the particle producing it was an electron. We use only single-event data, meaning that no cuts have been applied by looking for the subsequent neutron capture that would be seen in reactor antineutrino events. For Daya Bay (and KamLAND and PROSPECT, discussed below), we treat the signal detection efficiency as unity during the experiment's specified effective exposure, as the efficiency during livetime is very high. In Fig.~\ref{fig:dbplot}, we show the Daya Bay spectrum along with the expected spectrum of DM-induced proton recoils for a DM mass of 1 MeV and a range of cross sections.

We consider a particular DM mass and cross section to be ruled out if the predicted DM-induced recoil spectrum is higher than the total measured data at the 90\% confidence level at any energy. That is to say, if the predicted DM-induced flux is greater than the measured data at any energy and has less than a 10\% probability to fluctuate down to a level equal to or below the measured data, that mass-cross section pair is excluded. Note that this is different than the approach used by Ref.~\cite{Bri18}. That paper compared the total event rate and DM-induced recoil above 35 MeV proton kinetic energy, rather than the bin-by-bin comparison we make. For cross sections where attenuation is negligible, our method is more sensitive than comparing to the total Daya Bay event rate, because the shapes of the measured spectra and DM-induced spectra differ significantly. 

Figure~\ref{fig:dbplot} shows Daya Bay data along with predicted DM-induced recoil spectra for a DM mass of 1 MeV and two different cross sections, at the floor and ceiling of our exclusion region. The resulting exclusion region is shown, along with the exclusion region from the KamLAND data, in Fig.~\ref{fig:protonlimit}. \textit{This exclusion region almost completely closes the gap between the MiniBooNE and XENON1T limits}.  

\subsection{New Limits from KamLAND Data}

KamLAND is located approximately 1 km underground (2700 m.w.e.), comparable to both XENON1T and Borexino. The mountain above KamLAND is wider than it is high, with the main access tunnel running 1.7 km to one side of the mountain and an additional train tunnel running 3 km from KamLAND to another side of the mountain. For this reason, and because KamLAND is surrounded by other mountains, we conservatively consider only DM coming from at least 15$\degree$ above horizontal, and assume that all of the DM we do consider passes through 2 km of rock. As for Daya Bay, we neglect attenuation in the atmosphere.

\begin{figure}[t]
\centering
\includegraphics[width=\columnwidth]{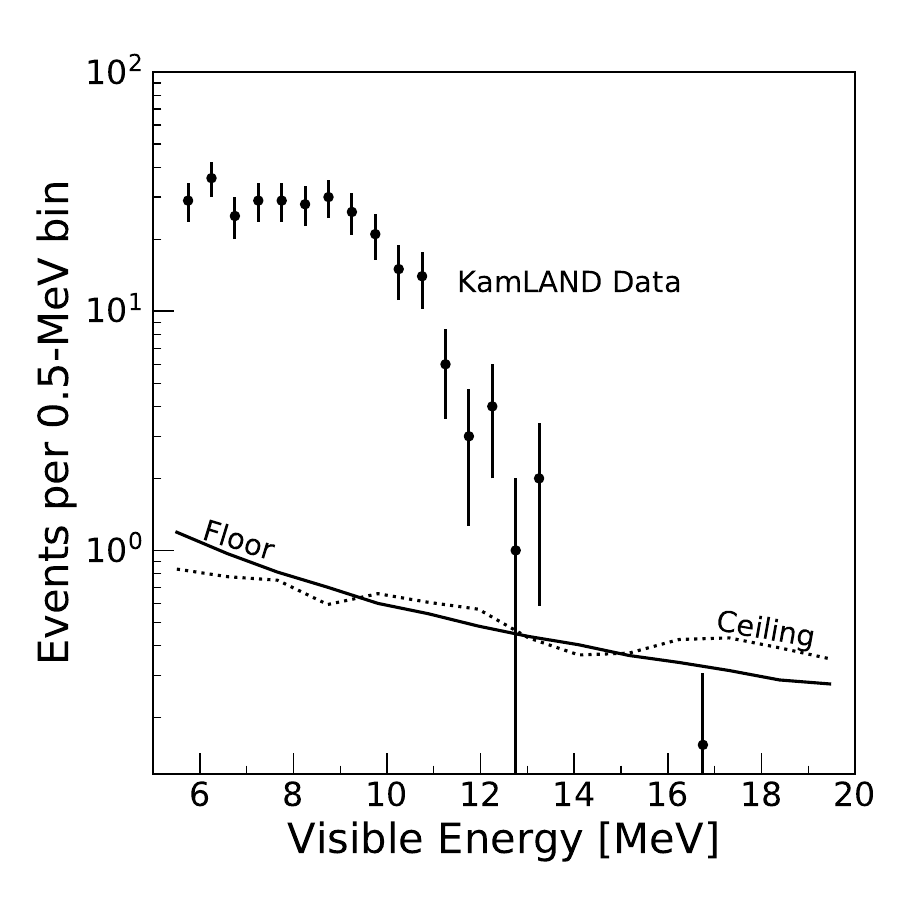}
\caption{KamLAND measured solar neutrino data \cite{Abe11} compared to predicted DM-induced spectra. Data from a 123 kton--day KamLAND observation (black points). In the 13.5--20 MeV range, KamLAND observed only one event. Also shown are two predicted DM-induced spectra for a mass of 1 MeV and cross sections of $3 \times 10^{-31}\, \rm{cm}^2$ and $1.3 \times 10^{-28}\, \rm{cm}^2$, corresponding respectively to the floor and ceiling of the KamLAND exclusion region in Fig.~\ref{fig:protonlimit}.} 
\label{fig:klplot}
\vspace{-0.25 cm}
\end{figure}

For KamLAND, we use data from a $^8$B solar-neutrino search \cite{Abe11}, which reports events from 5.5--20 MeVee of visible energy. The signal for a solar-neutrino search is elastic neutrino-electron scattering, unlike a reactor antineutrino search, which looks for a positron followed by a neutron capture. We consider the full spectrum, which includes contributions from other backgrounds.

\begin{figure}[t]
\centering
\includegraphics[width=\columnwidth]{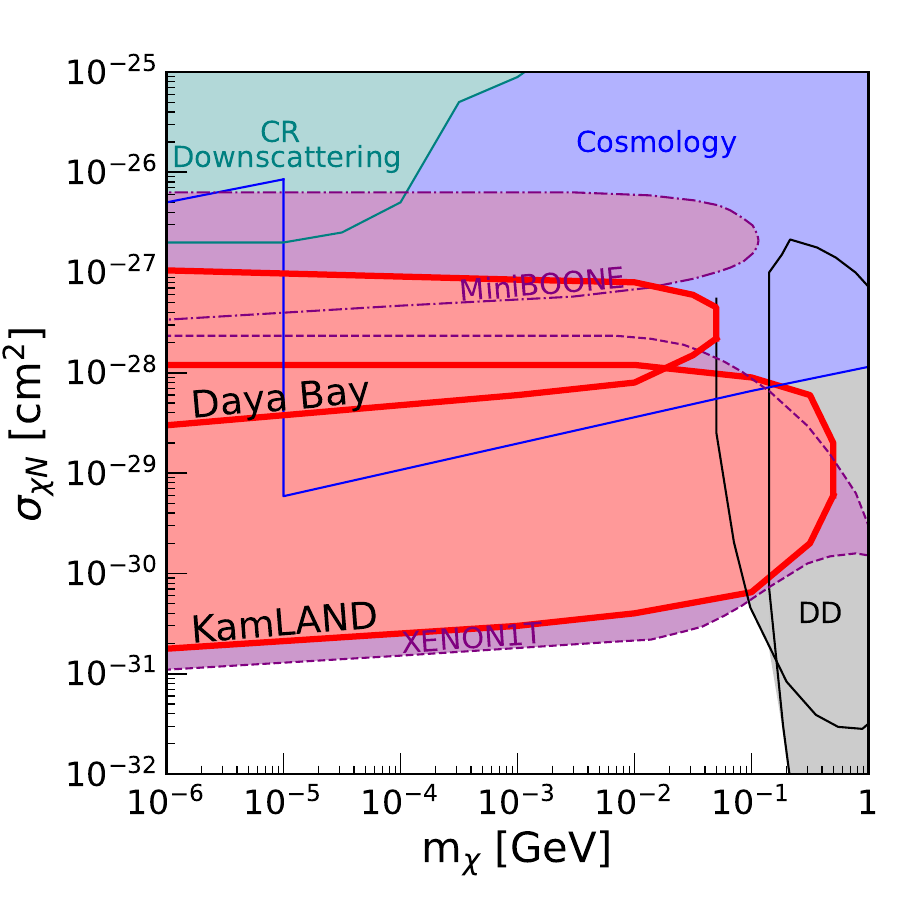}
\caption{Our exclusion regions calculated from KamLAND and Daya Bay data (red, labeled), compared to previous results from Ref.~\cite{Bri18} (purple), as well as existing limits from direct detection \cite{Col18, Abd19} (gray), cosmology\cite{Nad19, Glu18, Sla18, Xu18} (blue), and CRs \cite{Cap19} (teal). DM lighter than $\sim$1 keV, denoted by the vertical dashed line, cannot have been in thermal contact with ordinary matter in the early universe, due to effects on structure formation (see Ref.~\cite{Pet12} and references therein).} 
\label{fig:protonlimit}
\vspace{-0.25 cm}
\end{figure}

Figure~\ref{fig:klplot} shows the $^8$B data along with the DM-induced event spectra for a mass of 1 MeV. We proceed as for the Daya Bay data, considering a DM cross section to be ruled out for a given mass if the DM-induced event rate in the detector is significantly larger than the observed event rate in any energy bin. The resulting KamLAND exclusion region is shown in Fig.~\ref{fig:protonlimit}. \textit{Between KamLAND and Daya Bay, we exclude almost the entire XENON1T region through direct scattering on protons}.

\subsection{Sensitivity for PROSPECT and JUNO}

The PROSPECT reactor neutrino experiment \cite{Ash18} is located on Earth's surface at Oak Ridge National Laboratory. Although its background rate is high, its minimal shielding, less than even MiniBooNE's overburden, makes it ideal for probing large cross sections. PROSPECT is shielded by a meter of hydrogenous material, half a meter of concrete, and the atmosphere, of which the atmosphere is the dominant contribution. Because the atmospheric column density is at least 10 m.w.e.\ and because, unlike for neutron scattering, protons are not especially effective shielding compared to other nuclei, we neglect the hydrogenous material. We model the concrete as rock, and because it is a subdominant contribution to the shielding, uncertainties in its composition represent less than a $\sim$10\% effect.

\begin{figure}[t]
\centering
\includegraphics[width=\columnwidth]{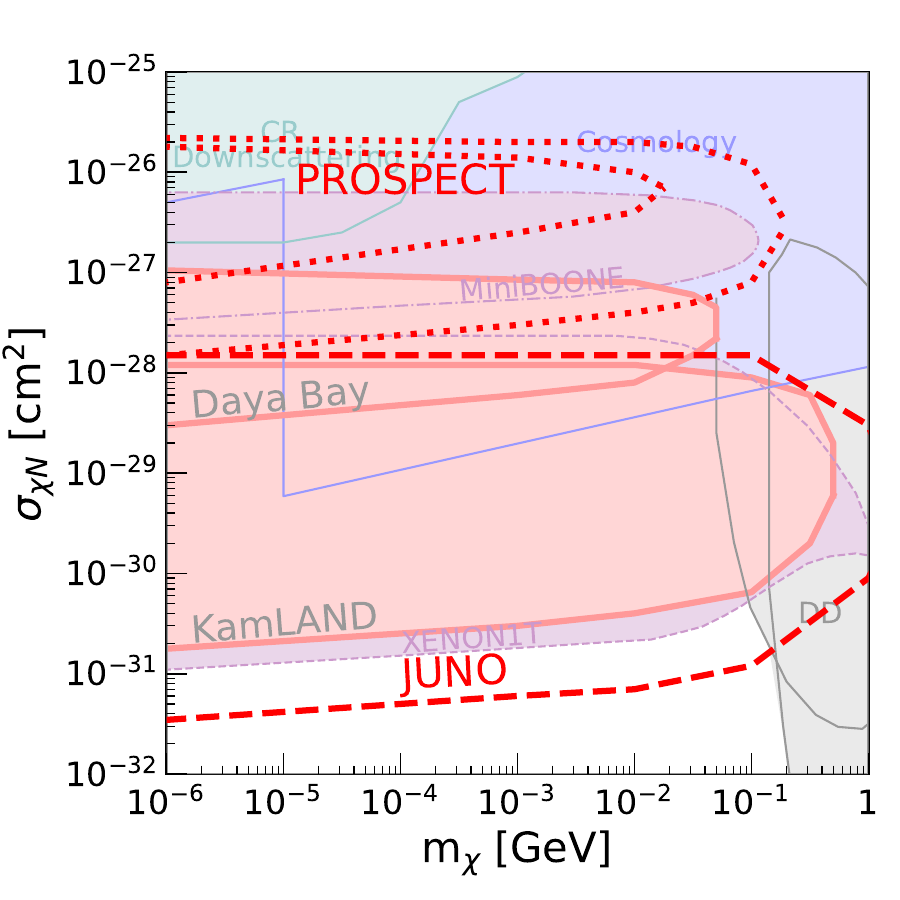}
\caption{Projected sensitivity regions for PROSPECT (red, dotted) and JUNO (red, dashed), compared to existing exclusion regions. See text for the difference between the two PROSPECT projections.} 
\label{fig:projections}
\vspace{-0.25 cm}
\end{figure}

We use the reactor-off spectrum of electron-like events in the energy range 0.8--8.8 MeVee from Ref.~\cite{Ash18}. In the available data, the background rate is too high to set a limit using the same analysis as above. However, the vast majority of the events represented by these data are expected to be electron, not proton, recoils. Applying pulse shape discrimination (PSD) cuts to the data could reduce the event rate by a factor of $\sim$1000 \cite{privcomm}. We derive a projected sensitivity based on the available data with an assumed factor of 1000 background reduction. Similar to above, we exclude a DM mass and cross section pair if the resulting CR-induced event rate is greater than the total PSD-reduced event rate at the 90\% level in any bin. The resulting sensitivity is shown in Fig.~\ref{fig:projections}. This region covers the largest cross sections probed by any direct-detection experiment for $m_{\chi} \ll 1$ GeV.

Additionally, the sensitivity could be further improved with precise background modeling. We have adopted the very conservative approach of comparing the DM-induced event rate to the total event rate in a given bin, but there are known to be standard model backgrounds for PROSPECT (and other detectors). Perfect background modeling would allow us to improve sensitivity by comparing the DM-induced event rate not to the total background, but to the statistical uncertainty on the background. This sensitivity, derived as a 90$\%$ CL and conservatively assuming one week of reactor-off data, is also shown in Fig.~\ref{fig:projections}. The true sensitivity, including systematic as well as statistical uncertainty in background modeling, would lie somewhere in between.

Finally, JUNO is a future reactor neutrino experiment currently under construction, at a depth of about 700 meters and with much larger volume than KamLAND \cite{Dju15}. Making the same assumption of perfect background modeling as above, we can compute the sensitivity of JUNO for a one year exposure. We consider only visible energies of 13.5 -- 20 MeV, the same energy bin which set our KamLAND limit, and assume that the background rate is the same as that measured by KamLAND in Refs.~\cite{Abe11, Gan11}, which is dominated by atmospheric neutrinos. In fact, the muon-induced neutron background should be much higher, but should only increase the total background by a factor of $\sim 2$. JUNO will also self-shield against neutrons, meaning that a slightly smaller fiducial volume could be used to largely remove this background. The derived 90$\%$ CL sensitivity region is shown in Fig.~\ref{fig:projections}. With only a year of exposure, \textit{JUNO can probe the lowest cross sections ever reached by direct detection for a large range of sub-GeV masses}.

\subsection{Astrophysical Uncertainties}

There is some uncertainty in our results in the local DM density, which is unavoidable for any direct detection experiment. The local DM density is typically taken to be 0.2--0.4 GeV/cm$^3$, and we use a standard value of 0.3 GeV/cm$^2$, so we take the uncertainty on the local DM density to be ${\sim}33\%$. There is also uncertainty in the steepness of the galactic DM density profile, especially close to the galactic center. Models such as the Burkert profile \cite{Bur95} have much lower DM density near the galactic center than the NFW profile that we employ. To quantify this uncertainty, we write down an extreme model in which the density profile follows an NFW shape for $r > 8$ kpc (the distance from Earth to the galactic center), but is held constant for $r < 8$ kpc. We find that the flux of upscattered dark matter from this cored profile is ${\sim}25\%$ lower than for the NFW profile. Adding these uncertainties in quadrature, we find a ${\sim}40\%$ uncertainty on the DM flux due to uncertainties in the DM density profile. Because our limits scale with the square root of the dark matter flux, this would result in only a ${\sim}20\%$ uncertainty on our limits.

For CR inputs, we make strictly conservative choices. As mentioned above, there is evidence that the CR flux is higher closer to the galactic center, which would make our results somewhat stronger. The LIS from Refs.~\cite{Bos17, Bos18a} are somewhat lower than the data from Voyager measurements shown in the same papers. And we assume a conservative size for the CR halo, and considering a larger volume would improve our results. 

But the uncertainties associated with the DM and CR densities are small compared to the statistical uncertainty on the background, and originate from our choice to compare the DM-induced event rate with the total background. For KamLAND, for example, there is one observed event in the considered energy range, but due to the large statistical uncertainty associated with a single event, we require four DM-induced events. While astrophysical inputs produce a $\sim20\%$ uncertainty on our limit, the low statistics produce a factor of 2 uncertainty. For Daya Bay, the effect of treating all observed events as potentially due to DM, rather than performing any background subtraction, is similarly large compared to astrophysical uncertainties.

\subsection{Future Ways to Improve Sensitivity}

It is apparent from Fig.~\ref{fig:protonlimit} that detectors at different depths cover different regions of parameter space, but share a similar shape: the ceiling of each region is set by attenuation in the atmosphere and Earth, the floor is set by detector backgrounds, and the high-mass end of each region is set by how they join. Because detectors at different depths are sensitive to different ranges of cross section, considering additional detectors may provide useful new coverage. We have considered only two of the Daya Bay ADs in setting our exclusion regions, and considering the other (deeper) detectors could push Daya Bay's sensitivity to lower cross sections. Probing cross sections above the PROSPECT sensitivity region would require detectors with minimal atmospheric shielding, such as \mbox{balloon-,} \mbox{rocket-,} or satellite-based experiments.

\section{Constraining the DM-Electron Cross Section}

In this section, we apply the general treatment presented in Sec.~\ref{sec:recoildist} to DM-electron scattering. For our new results, we use lower-energy data than Ref.~\cite{Ema18}, and show that doing so yields a larger signal to background ratio at a given DM mass and cross section, thus producing stronger limits.

\subsection{Electron Recoil Distribution}

The case of DM-electron scattering is simpler than DM-nucleon scattering because we consider only one CR species, and because there is no form-factor suppression. Additionally, we do not have to convert between nuclear recoil energy and electron recoil energy in the detector. So we obtain the DM flux directly from Eqs.~(\ref{DMdist}) and (\ref{DMdist2}), with the DM-electron cross section $\sigma_{\chi e}$ being the relevant cross section.

In computing the DM flux at Super-K, we consider only DM arriving from at least 15$\degree$ above horizontal, and neglect attenuation for this flux. Over the entire DM mass range we consider, the lowest cross section probed by Ref.~\cite{Ema18} is orders of magnitude below the ceiling caused by attenuation in Earth. We restrict our attention to cross sections too small to be excluded by Ref.~\cite{Ema18}, so attenuation is negligible. As a result, we can insert the computed DM spectrum directly into Eq.~(\ref{recoil}) to get the recoil spectrum per target electron, and multiply this by $7.5 \times 10^{33}$ electrons \cite{Ema18} to get the recoil spectrum in Super-K.

We note, however, that the ceiling of our analysis could be lower if $\sigma_{\chi N}$ or the DM-proton cross section $\sigma_{\chi p}$ is closely related to $\sigma_{\chi e}$, such that increasing $\sigma_{\chi e}$ increases attenuation by nuclei as well (see, e.g., Ref.~\cite{Emk17b}).

\subsection{Our Improvements Over Previous Work}

\begin{figure}[t]
\centering
\includegraphics[width=\columnwidth]{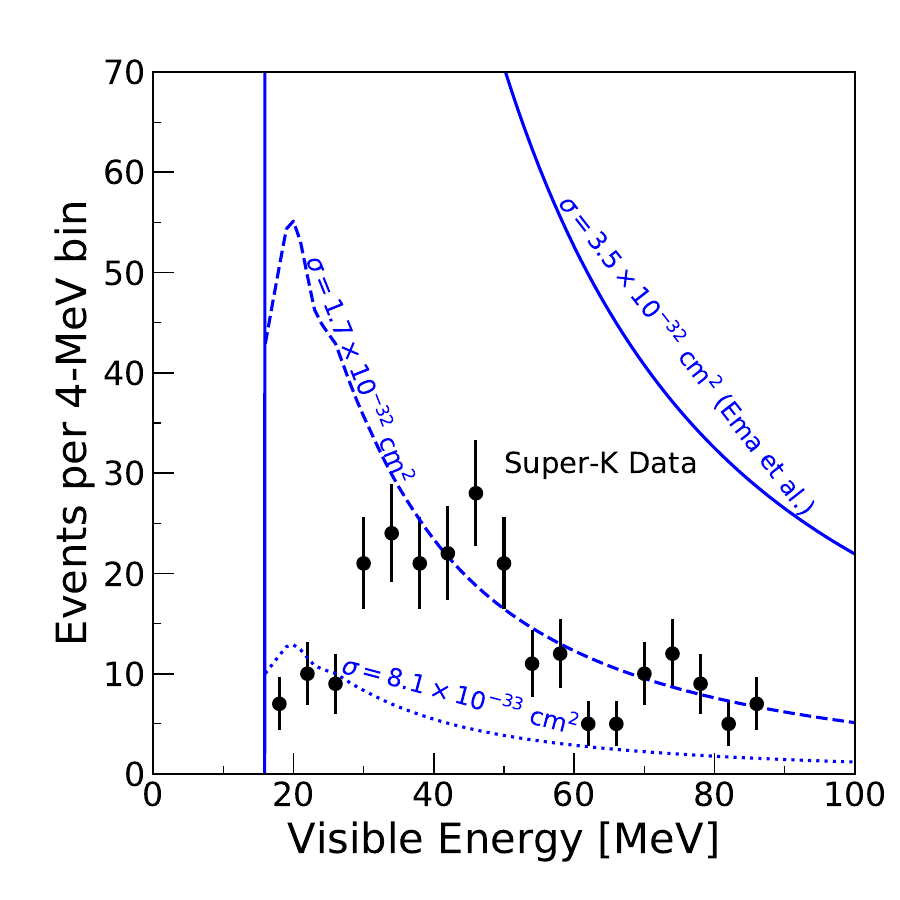}
\caption{Super-K measured data from a DSNB search, compared to predicted DM-induced spectra. Data from a 1497--day Super-K observation, from 16--88 MeV \cite{Bay11} (black points). Also shown are three predicted DM-induced recoil spectra for a mass of 1 MeV. The increasingly large cross sections correspond to our limit, the middle of the new region we exclude, and the previous limit from Ref.~\cite{Ema18}. The wiggles in the spectra at low energy are due to signal detection efficiency, as is the vertical blue line, below which we take efficiency to be zero (see text).} 
\label{fig:skspectrum}
\vspace{-0.25 cm}
\end{figure}

As seen in our Fig.~\ref{fig:skspectrum}, the DM-induced recoil spectrum is steeply falling, meaning that for a given cross section, a higher signal rate is observed at lower energy. Below, we show that considering data below 100 MeV produces significantly stronger limits than previously derived. We also derive projected limits from Hyper-Kamiokande that are stronger than the previously derived projections for Super-K and DUNE.

\subsection{New Limits from Super-K Data}

In the previous analysis, Ref.~\cite{Ema18} considered Super-K data above 100 MeV in order to stay above the atmospheric neutrino background. We instead consider energy as low as 10 MeV, a range where the atmospheric neutrino background falls with decreasing energy while the predicted DM spectrum rises sharply, in order to obtain a higher signal-to-background ratio. We consider the spectrum of electron-recoil events from a diffuse supernova neutrino background (DSNB) search performed using Super-K data from the SK-I exposure, from April 1996 to July 2001 \cite{Bay11}. During this time period, 239 electron-recoil events are reported in the energy range from 16--88 MeV visible energy (related to our work, these data have also been used to set constraints on dark radiation \cite{Cui17}). Note that because we consider electron-recoil events, there is no need to include quenching, as was done in the previous section to convert nuclear recoil energy to electron-equivalent energy. We follow the same approach as for DM-nucleon scattering, considering a DM mass and cross section ruled out if the predicted DM-induced event rate is higher than the measured event rate at the 90\% level at any energy. 

Some fraction of DM-induced events may be lost to analysis cuts. We obtain the signal detection efficiency as a function of recoil energy by interpolating Fig. 10 of Ref.~\cite{Bay11}, and multiply our computed event spectrum by the efficiency in order to get the spectrum of DM-induced events passing analysis cuts. As the data only extend down to 16 MeV, we take the signal efficiency to be zero below this energy.

Figure~\ref{fig:skspectrum} shows the Super-K data along with the (efficiency-corrected) spectrum of CR-induced DM events for a mass of 1 MeV and several values of the cross section. The increasingly large cross sections, as labeled, correspond to our limit, the halfway point between our limit and the previous limit from Ref.~\cite{Ema18}, and the limit from Ref.~\cite{Ema18} itself. The resulting limit is shown in Fig.~\ref{fig:sklimit}, along with the limits from Ref.~\cite{Ema18}, as well as astrophysical, cosmological, direct detection and solar reflection limits.

Our limit is tighter than the limit set by Ref.~\cite{Ema18} by a factor of at least $\sim 4$, depending on the mass. It is comparable to the Super-K sensitivity curve derived in Ref~\cite{Ema18}, and close to their DUNE sensitivity curve. This means that \textit{we have already excluded most of the parameter space which was previously thought to be reachable only with careful background subtraction, restriction to events coming from the galactic center, and possibly several years of DUNE data}. 

\subsection{Sensitivity for Hyper-Kamiokande}

\begin{figure}[t]
\centering
\includegraphics[width=\columnwidth]{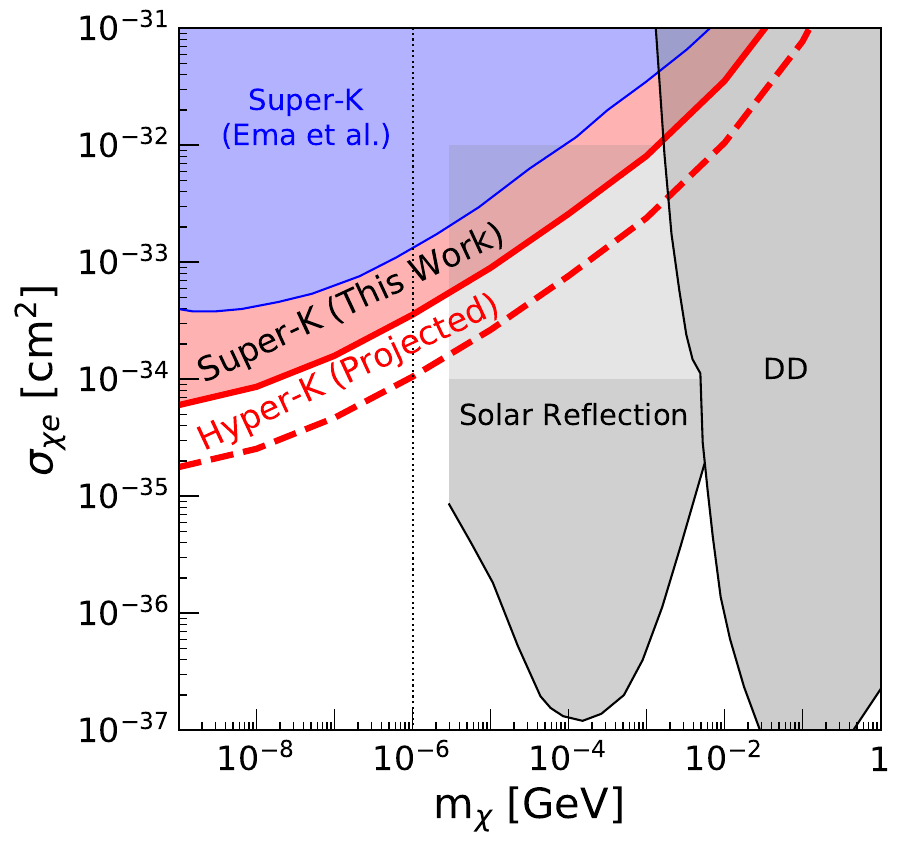}
\caption{Our extension of the exclusion region calculated from Super-K data (solid red) and our projection for Hyper-K (dashed red) compared to limit from Ref.~\cite{Ema18} (blue), as well as previous limits from direct detection (DD) \cite{Abr19, Agn18, Agn18D, Ess17} and solar reflection (dark grey region from Ref.~\cite{An18}; Ref.~\cite{Ema18} argued that the ceiling of their region should be higher, and this extension is shown by the light gray box above it). The limit based on CR spectra \cite{Cap19} constrains cross sections higher than shown on this plot. DM lighter than $\sim$1 keV, denoted by the vertical dashed line, cannot have been in thermal contact with ordinary matter in the early universe, due to effects on structure formation (see Ref.~\cite{Pet12} and references therein).} 
\label{fig:sklimit}
\vspace{-0.25 cm}
\end{figure}

Hyper-Kamiokande is a proposed neutrino experiment that will consist of two large water Cherenkov detectors, with a total fiducial volume 16.8 times that of Super-Kamiokande \cite{Abe16}. Assuming that the observed background will be similar to that of Super-Kamiokande, we can compute the 90$\%$ CL sensitivity in the same way as we did for PROSPECT and JUNO. We assume an exposure 16.8 times larger than the Super-Kamiokande exposure used to set our limit. This projection, shown in Fig.~\ref{fig:sklimit}, is stronger than previous projections for Super-K and DUNE and at least an order of magnitude stronger than previous constraints.

\subsection{Future Ways to Improve Sensitivity}

Unlike in the case of DM-nucleon scattering, we do not consider detectors at different depths for DM-electron scattering, nor would doing so improve on existing limits. Super-K is already very deep, and higher cross sections are already covered: though it is not shown in Fig.~\ref{fig:sklimit}, Ref.~\cite{Ema18} also derived a constraint based on MiniBooNE data, which covers cross sections above the ceiling they computed for the Super-K region. Still larger cross sections are constrained by considering CR downscattering \cite{Cap19} and cosmological observations \cite{Ali15}. We refer the reader to Ref.~\cite{Ema18} for a plot of the higher cross section constraints.

Additionally, in a detector that has directional sensitivity, as Super-K does, it is possible to search for events coming only from the direction of the galactic center. Ref.~\cite{Car15} has studied the morphology of a hypothetical signal coming from CR-DM interactions, and the signal distribution they compute could be useful for such a search. These possibilities have both been discussed in Ref.~\cite{Ema18}.

A coming improvement to the Super-K detector is the addition of water-soluble gadolinium salt, which will allow tagging of antineutrino events, greatly reducing backgrounds \cite{Bea03, Lab18}. Another is the further reduction of spallation backgrounds, which are the dominant background in Super-K from 6--18 MeV, based on the cuts proposed by Refs.~\cite{Li14, Li15a, Li15b}. Background reduction at low energy, where the predicted DM-induced event rate is largest, will improve sensitivity.

\section{Directions for Future Work}\label{sec:discussion}

Here we mention new directions for testing low-mass DM, beyond the discussions above about improving the sensitivity of this paper's results.

We first note that the energy dependence of the cross sections may be non-constant, contrary to the energy-independent case we assume, which would change the exclusion regions. If the cross sections increase with energy, as is likely, this would improve the sensitivity, although it may also lower the ceiling of the corresponding region. See Ref.~\cite{Dent19, Bon19} for initial explorations. More generally, it will be interesting to develop concrete models for these light DM scenarios; that could also provide connections to constraints from fixed-target experiments \cite{Dia13, Bat14, Iza14,  Kah14, Sop14, Col15, deN16, Ban17, Agu18}, as done in Ref.~\cite{Bon19}.

We have only considered elastic scattering on protons and electrons, but other target nuclei and additional signals could also be considered. Carbon recoils in liquid scintillator would be heavily quenched, but would benefit from coherent enhancement of the cross section, possibly allowing neutrino detectors to probe lower cross sections. Carbon nuclei in a scintillator-based detector could also be excited by a collision with high-energy DM, and the de-excitation would produce a characteristic 15.11 MeV gamma-ray signal. Additional signals such as bremsstrahlung photons, considered by Refs.~\cite{Kou16, McC17, Ake18, Bel19, Liu19}, and the Migdal effect, as discussed by Refs.~\cite{Ibe17, Dol17, Ake18, Arm19, Bel19, Liu19}, could be seen in low-threshold detectors. 

A DM particle with sufficient energy could also perhaps expel a neutron from a carbon nucleus. The resulting neutron capture and subsequent decay of the unstable $^{11}$C nucleus would be a distinctive delayed-coincidence signal \cite{Gal04}. An analogous delayed coincidence signal could be seen in Super-K, with a neutron expelled from an oxygen nucleus, leading to both a neutron capture signal and the subsequent gamma-ray deexcitation of an excited $^{15}$O nucleus \cite{Lan95}. For even higher-energy DM, collisions with nuclei could produce pions, and the subsequent pion decays could be additional signals to search for. Future work could also consider inelastic and quasi-elastic contributions to the attenuation, although we expect these effects to be subdominant in the energy range that drives our limits.

\section{Conclusions}

We have expanded and improved upon previous studies of cosmic ray-upscattered dark matter, in which dark matter that is nominally too light to be detected is accelerated by cosmic rays to detectable energies. We have developed the first numerical propagation code to model attenuation of light, relativistic dark matter, resulting in more accurate ceilings which allow us to probe higher cross sections than the previously used ballistic approximation. We have excluded parameter space for dark matter-proton scattering never before probed by direct detection, and identified detectors that can expand this coverage even further. We have also substantially increased the parameter space probed directly by dark matter-proton scattering, making our results complementary to xenon-based limits. Finally, we have improved the sensitivity of Super-K to dark matter-electron scattering, for some masses superseding even the future projections derived in previous work, and derived the sensitivity of the future Hyper-Kamiokande detectors. 

%Though direct-detection experiments are extremely sensitive to GeV-scale dark matter, they dramatically lose sensitivity below about a GeV. Dark matter is typically assumed to be at most weakly interacting, but the tightest bounds on sub-GeV dark matter are more than 15 orders of magnitude weaker than the limits on GeV-scale particles. Additional probes are necessary in order to test light dark matter with the same sensitivity. One such probe is cosmic ray-dark matter scattering: such collisions would both alter the observed cosmic ray spectra and boost light dark matter to high energy.

%Here we have focused on the latter effect, and shown that a variety of neutrino experiments may be sensitive to sub-GeV dark matter elastically scattering with nuclei or electrons. Although such light dark matter is typically below detector thresholds, collisions with cosmic rays can upscatter it to much higher energy, making it detectable. More significantly, we have shown that considering different detector depths and energy ranges is crucial for optimizing this probe's sensitivity to a range of dark matter cross sections. 

Though our limits are based on the astrophysics of cosmic rays and the galaxy's dark matter halo, our results are fairly insensitive to astrophysical uncertainties. Unlike traditional direct-detection searches, the observed event rate scales with the square of the cross section, so that our limits depend only on the square root of the cosmic ray flux or dark matter density. Uncertainties in the interstellar cosmic ray spectra, cosmic ray halo size, and dark matter profile are suppressed by this square root and thus largely negligible.

Future work could lead to significant improvement on our results, extending sensitivity to both higher and lower cross sections. Significant background reduction and precise background modeling would lead to strengthened constraints on both dark matter-proton and dark matter-electron scattering. With such developments, as well as consideration of specific dark matter models, cosmic rays will only become more powerful as a probe of sub-GeV dark matter, constraining the parameter space between collider and cosmological limits.

\bigskip

\section*{Acknowledgments}
We are grateful to the anonymous referee for insightful comments and suggestions, which substantially improved the paper. We are grateful for helpful discussions with Nathaniel Bowden, Jun Cao, Eric Carlson, Timon Emken, Rouven Essig, Glennys Farrar, Karsten Heeger, Thomas Langford, Bryce Littlejohn, M. Shafi Mahdawi, Pieter Mumm, Kenny Ng, Annika Peter, Pranava Teja Surukuchi, Lindley Winslow, and Xingchen Xu, and especially Torsten Bringmann, Yohei Ema, Maxim Pospelov and Filippo Sala.

CVC and JFB are supported by NSF grant PHY-1714479.

\bibliography{BoostedDM3}

\pagebreak

\onecolumngrid

\begin{center}
{\large\bf Erratum: Strong New Limits on Light Dark Matter from Neutrino Experiments}\\
{\large\bf [Phys. Rev. D 100, 103001 (2019)]}

\bigskip

{\bf Christopher V. Cappiello}\\
{\tt cappiello.7@osu.edu}
\medskip

{\bf John F. Beacom}\\
{\tt beacom.7@osu.edu}

\bigskip

(Erratum Date: \today)
\end{center}

Due to a numerical error in our propagation code, the nuclear form factor was not properly accounted for when modeling propagation of dark matter (DM) through the Earth's crust. This same issue also affected the interaction of DM particles with target protons within a detector. That error, and the changes described here, affect only our treatment of DM-nucleon scattering and not DM-electron scattering. The most noticeable effect of fixing these errors is to reduce attenuation. So at large cross sections, our limits extend higher by a factor of up to 3.0. At small cross sections, our limits get weaker, but by a negligible amount.  Here we revise Figs.~\ref{fig:dbatten}--\ref{fig:protonlimit}. Fig.~\ref{fig:projections} is also affected, but as it only shows future projections and the effect of these changes is small, we do not reproduce it.

%Figures~\ref{fig:dbattene}, \ref{fig:dbplote}, and \ref{fig:klplote} show DM-induced event spectra for Daya Bay and KamLAND for several DM cross sections. The cross sections presented in Fig.~\ref{fig:dbattene} are unchanged, and all that changes are the spectra associated with these cross sections. In Figs.~\ref{fig:dbplote} and \ref{fig:klplote}, we plot spectra corresponding to cross sections at the top and bottom of our exclusion regions. As our exclusion regions have changed somewhat, we remade these figures with new cross sections. These new cross sections are specified in the figure captions. Finally, Fig.~\ref{fig:protonlimite} shows our updated limit on DM-nucleon scattering.

%In summary, our updated limits on DM-nucleon scattering extend to somewhat larger cross sections than before. Our results for DM-electron scattering are unchanged.

\medskip

We are grateful to Kazumi Hata for pointing out one of these errors.

\begin{figure}[h]
\centering
\includegraphics[width=0.5\columnwidth]{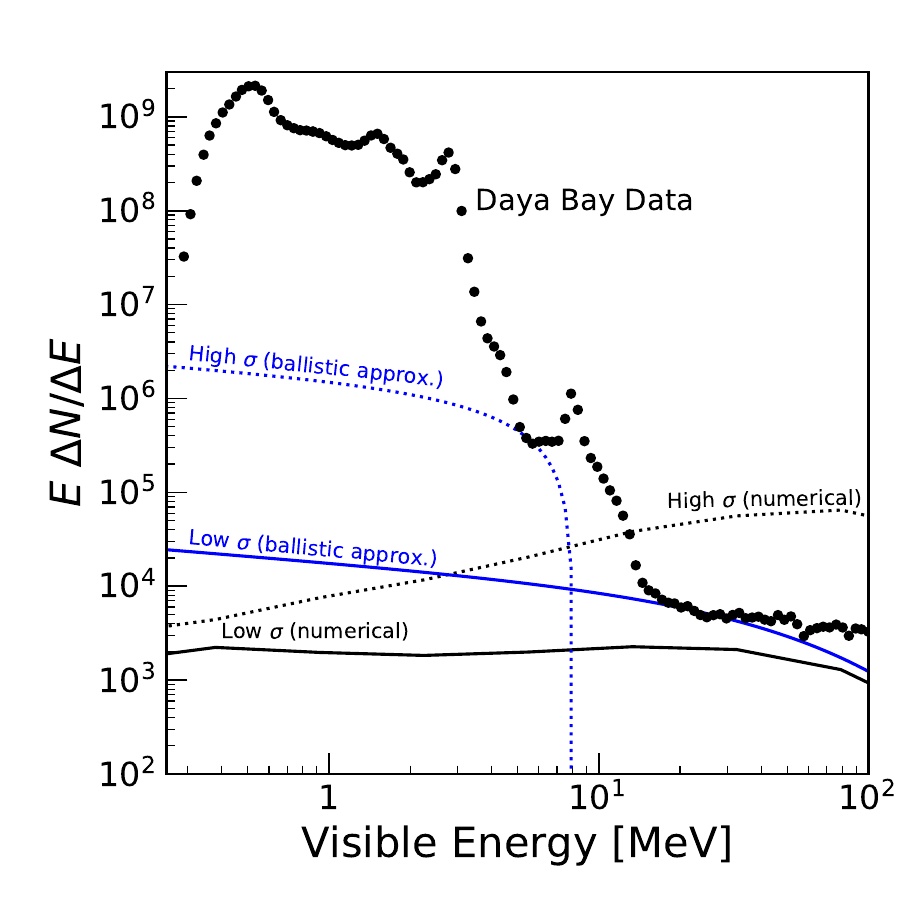}
\caption{Updated version of Fig.~\ref{fig:dbatten}, using the same cross sections as in the main text. The ``High $\sigma$ (numerical)" curve is significantly higher than before, while the ``Low $\sigma$ (numerical)" curve is slightly lower.} 
\vspace{-0.25 cm}
\label{fig:dbattene}
\end{figure}

\begin{figure}[h]
\centering
\includegraphics[width=0.5\columnwidth]{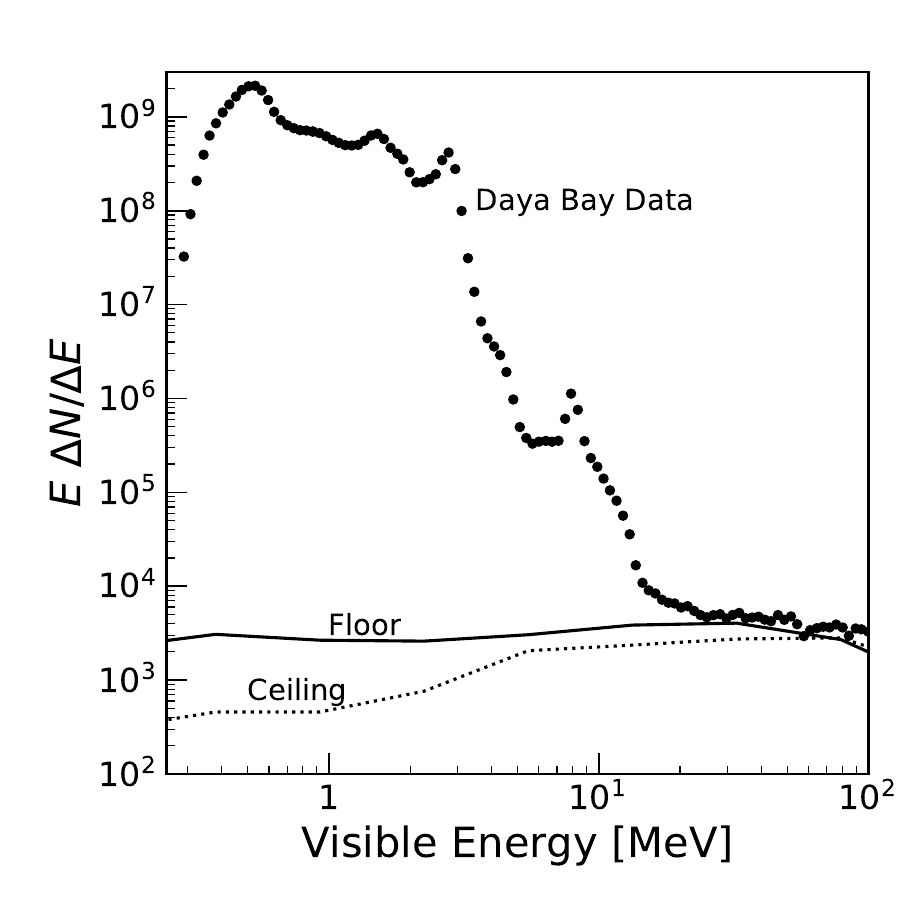}
\caption{Updated version of Fig.~\ref{fig:dbplot}. The ``floor" curve now corresponds to $8.0 \times 10^{-29}$ cm$^2$ instead of $6.0 \times 10^{-29}$ cm$^2$. The ``ceiling" curve now corresponds to $2.3 \times 10^{-27}$ cm$^2$ instead of $7.0 \times 10^{-28}$ cm$^2$. These curves are qualitatively similar to the previous curves, because by definition the ceiling and floor are just barely ruled out by the data. This means that fixing the errors in the form factor, and subsequently changing the cross section to correspond to the new floor or ceiling, has the effect of producing similar spectra to those shown previously.} 
\vspace{-0.25 cm}
\label{fig:dbplote}
\end{figure}

\begin{figure}[h]
\centering
\includegraphics[width=0.5\columnwidth]{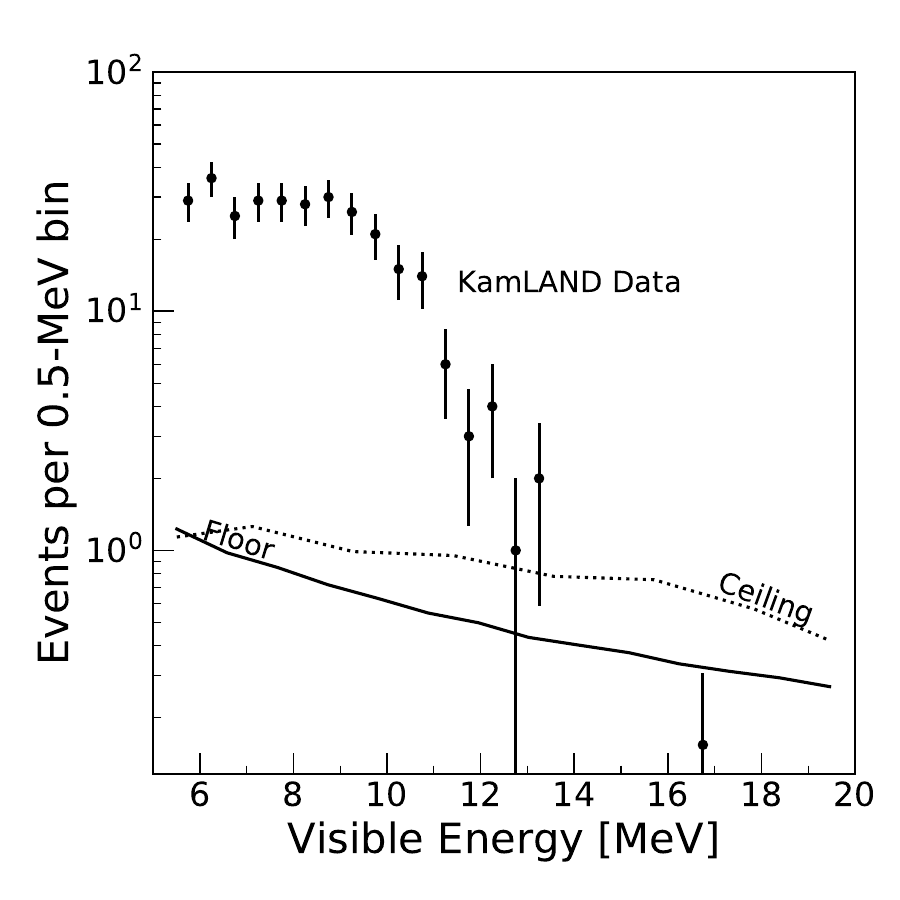}
\caption{Updated version of Fig.~\ref{fig:klplot}. The ``floor" curve now corresponds to $4.0 \times 10^{-31}$ cm$^2$ instead of $3.0 \times 10^{-31}$ cm$^2$. The ``ceiling" curve now corresponds to $1.5 \times 10^{-28}$ cm$^2$ instead of $1.3 \times 10^{-28}$ cm$^2$. These curves are qualitatively similar to the previous curves, for the same reason as above.} 
\vspace{-0.25 cm}
\label{fig:klplote}
\end{figure}

\begin{figure}[h]
\centering
\includegraphics[width=0.5\columnwidth]{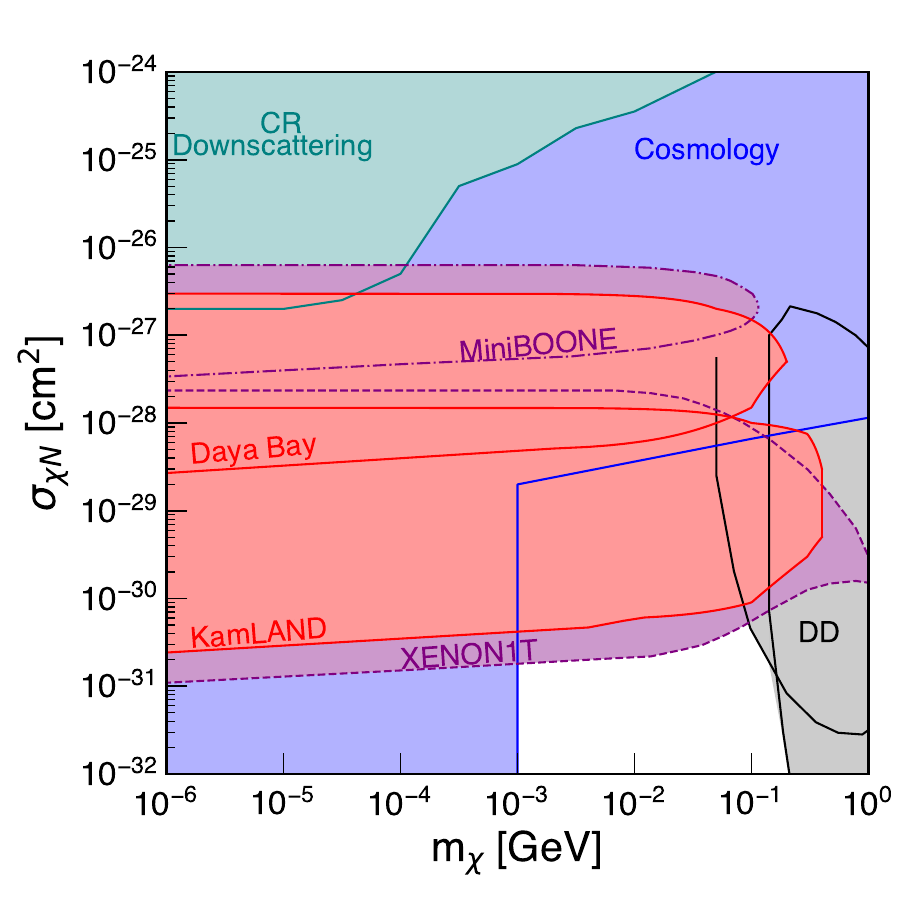}
\caption{Updated version of Fig.~\ref{fig:protonlimit}, showing constraints on DM-nucleon scattering. The main change is that the exclusion regions extend to somewhat larger cross sections. The exclusion region for Daya Bay also extends to noticeably larger masses.} 
\vspace{-0.25 cm}
\label{fig:protonlimite}
\end{figure}

\end{document}